\documentclass[twocolumn,twoside]{revtex4}
\usepackage[utf8]{inputenc}
\usepackage{amsmath}
\usepackage{amsfonts}
\usepackage{amssymb}
\usepackage{graphics}
\usepackage{graphicx}
\usepackage{subfigure}
\usepackage{pdfpages}
\usepackage{hyperref}
\usepackage{fancyhdr}
\usepackage{tikz}
\usepackage[compat=1.1.0]{tikz-feynman}
\usepackage{tikz-feynman}
\begin{document}

\title{Results and prospects of radiative and electroweak penguin decays at Belle II}

%

\author{Soumen Halder\\
(On behalf of the Belle II Collaboration)}
\email{soumen.halder@tifr.res.in}
\affiliation{Tata Institute of Fundamental Research, Mumbai 400005, India}
\begin{abstract}
The $b\to s(d)$ quark-level transitions are flavor-changing neutral current processes, which are not allowed at tree level in the standard model.
These processes are very rare and constitute a potential probe for new physics. 
Belle II at SuperKEKB is a substantial upgrade of the Belle experiment.
It aims to collect 50 ab$^{-1}$ of data with a design peak luminosity of $8\times 10^{35}$ cm$^{-2}$s$^{-1}$ that is 40 times more than its predecessor.
It has been recording data since 2019 and during these early days of the experiment, efforts are being made to  detect early signals of the above decays. 
We report the first reconstrution in Belle II data  of a $B\to K^{*}\gamma$ signal as well as future prospects for radiative and electroweak decays at Belle II.
\end{abstract}

\maketitle

\thispagestyle{fancy}

\section{Introduction}

\begin{figure}[h]
    \begin{tikzpicture}
	\begin{feynman}
	
	\vertex(c){ $b$};
	\vertex[right=3.5cm of c] (d){ $s$};
	\vertex[right=1cm of c] (e);
	\vertex[right=1.5cm of e] (f);
	\vertex[above=1cm of e](g);
	\vertex[above=1cm of f](h);
	\vertex at ($(g)!0.6!(h)!2cm!60:(h)$) (i){ $\ell^{+} (\nu_{\ell})$};
	\vertex at ($(g)!0.6!(h)!1.5cm!30:(h)$) (j){ $\ell^{-} (\nu_{\ell})$};
	
	\vertex [below=0.5em of e](z);
	
	\vertex [below=0.5em of f](z1);
		
	\diagram* {
		(c) --[fermion] (e),
		(e) --[fermion, edge label={ $t$}] (f),
		(f) --[fermion] (d),
		(e) -- [boson, edge label= {\small \(W^{-}\)}] (g),
		(h) -- [ boson, edge label= \(W^{-}\)] (f),
		(g) -- [ fermion, edge label=  $\nu_{\ell}(\ell)$] (h),
		(i) -- [ fermion] (g),
		(h) -- [ fermion] (j),
	};
	\end{feynman}	
	\end{tikzpicture}
		\begin{tikzpicture}
	\begin{feynman}
	
	\vertex(c){ $b$};
	\vertex[right=3.5cm of c] (d){ $s$};
	\vertex[right=1cm of c] (e);
	\vertex[right=1.5cm of e] (f);
	\vertex[above=1cm of e](g);
	\vertex[above=1cm of f](h);
	\vertex at ($(g)!0.6!(h)!2cm!60:(h)$) (i){ $\ell^{+}$};
	\vertex at ($(g)!0.6!(h)!1.5cm!30:(h)$) (j){ $\ell^{-}$};
	
	\diagram* {
		(c) --[fermion] (e),
		(e) --[fermion] (f),
		(f) --[fermion] (d),
		(e) -- [boson, edge label= \(H^{-}\)] (g),
		(h) -- [ boson, edge label= \(H^{-}\)] (f),
		(g) -- [ fermion] (h),
		(i) -- [ fermion] (g),
		(h) -- [ fermion] (j),
	};
	\end{feynman}	
	\end{tikzpicture}
	
	\caption{Feynman diagrams of $b\to s\ell^{+}\ell^{-}$ featuring a SM box diagram (left) and non-SM box diagram where  the $W$ bosons are replaced by some non-SM particles such as charged Higgs bosons (right).}
		 \label{fig:intro}
	\end{figure}
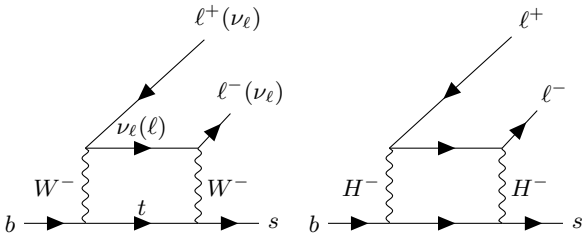

The flavor-changing neutral current processes mediated by $b\to s(d)$ transitions are forbidden at tree level in the standard model (SM).
These processes can however proceed via higher-order amplitudes involving quantum loops.
Non-SM particles may contribute in such loops as exemplified in Fig ~\ref{fig:intro}, which could suppress or enhance the amplitude of the decay rate.
Hence, the decays mediated by $b \to s (d)$ transitions potentially  probe new physics (NP).
In this article, we report the current status and future prospect of Belle II for radiative penguin decays proceeding via $b\to s(d)\gamma$ and for electroweak penguin decays mediated by $b\to s(d)\ell^{+}\ell^{-}$ or $b\to s(d)\nu \bar{\nu}$ transitions.
	
\section{SuperKEKB and Belle II}

SuperKEKB is the next generation $e^{+}e^{-}$ collider located at Tsukuba, Japan, which plans to collide $e^{+} $ and $e^{-}$ beams at a rate 40 times higher than its predecessor KEKB.
The Belle II detector placed at the collision point of SuperKEKB is a major upgrade of Belle.
It has collected about 0.5\,fb$^{-1}$ data during its pilot run in 2018, which was aimed at ensuring that beam background levels are safe to install the sensitive vertex detector.
Since the full detector integration in 2019, Belle II has recorded 55\,fb$^{-1}$ data.
The eventual goal is to collect 50\,ab$^{-1}$ of data, which will make the next decade very interesting for the flavor physics community.
A short summary on the Belle II experiment is available in Ref.~\cite{belle2}.

\section{Analysis techniques}
The analysis techniques used to study the rare decays can be divided into the following two categories. 
\begin{itemize}
    \item Exclusive: A specific $B$ meson decay mode is studied by reconstructing all of its final-state particles, for example, the analysis of the decay $B^{+}\to K^{+}e^{+}e^{-}$.
   \item Inclusive: In an inclusive analysis some of the final-state particles are not explicitly reconstructed.
   The study of $B\to X_{s}\gamma$ processes is an example of such inclusive analysis, where $X_{s}$ is defined as any final state having net strangeness of one.
   Inclusive decay analyses are further classified into two categories, namely semiinclusive and fully inclusive.
   Semiinclusive analyses are performed by combining several exclusive decay modes.
   Fully inclusive analyses do not rely on specific exclusive decays, rather they involve the reconstruction of the recoiling $B$ meson with the hadronic or semileptonic tagging procedure.
   A schematic diagram for these two types of inclusive analysis is shown in Fig.~\ref{fig:my_label0}.
   In the hadronic-tag inclusive analyses, the momentum of the signal $B$ meson is measured, whereas this is not feasible for the semileptonic-tag analyses due to the presence of a neutrino.
   Therefore, the former has a lower signal efficiency since it fully reconstructs tag-side $B$ meson from hadronic decays, which have relatively smaller branching fractions compared to semileptonic decays.
   On the other hand, the challenge of a semileptonic tag analysis lies in dealing with the relatively higher background level. 
\begin{figure}[h]
		\includegraphics[width=7cm]{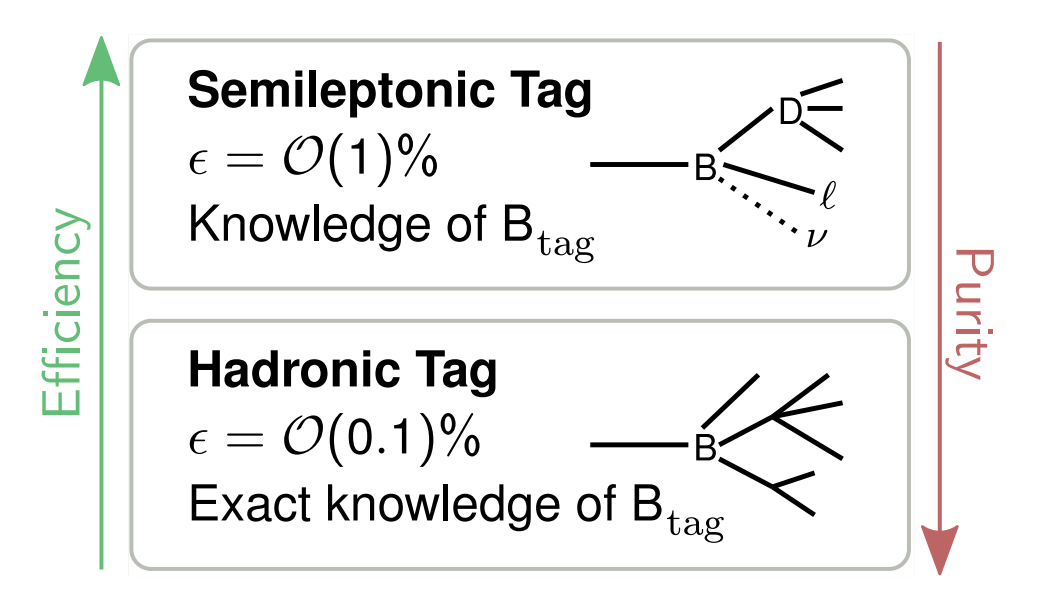}
		\includegraphics[width=8cm]{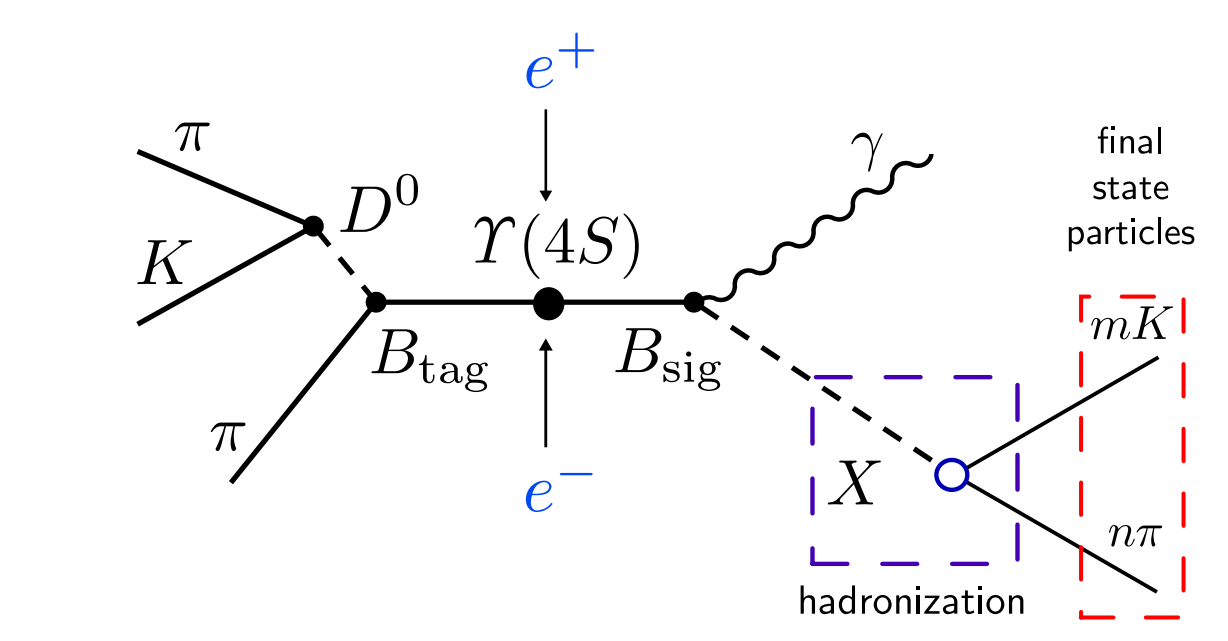}
    \caption{Comparison between semileptonic and hadronic tag in terms of purity and efficiency (top). Illustration of a hadronic-tagged $B\to X_{s}\gamma$ event in the center-of-mass frame (bottom). }
    \label{fig:my_label0}
\end{figure}
\end{itemize}

\section{Radiative penguin $B$ decays}

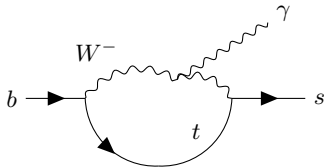
\begin{figure}[h]
	\begin{tikzpicture}
		\begin{feynman}
		\vertex (b1) { $b$};
		\vertex [right=3em of b1] (b2);.
		
		\vertex [right=6em of b2] (b3);
		\vertex [right=3em of b3] (b4) {$s$};
		
		\vertex at ($(b2)!0.4!(b3)!0.5cm!30:(b3)$) (g1);
		\vertex at ($(g1)!0.4!(b3)!1.5cm!60:(b3)$) (g2){$\gamma$};
		
		\vertex at ($(b2)!0.5!(b3)!0.9cm!-90:(b3)$) (x1);
		
		\diagram* {
			
			(b1) -- [fermion] (b2) -- [boson,quarter left, edge label={ $W^{-}$}] (g1) -- [boson,quarter left] (b3) -- [fermion] (b4),
			
			(b2) -- [fermion,quarter right] (x1) -- [plain,quarter right, edge label={ $t$}] (b3),
			(g1) -- [boson] (g2) 
			
		};
		\end{feynman}
		\end{tikzpicture}
	   \caption{Leading order Feynman diagram for the $b \to s\gamma$ process.}
    \label{fig:my_label9}
\end{figure}

In this section, we discuss $B$ decays that are mediated by $b\to s(d)\gamma$ transitions.
The leading order Feynman diagram for this process is shown in Fig.~\ref{fig:my_label9}.

\subsection{First reconstrution of the penguin $B$ decay in Belle II data}

Among the radiative penguin decays, $B\to K^{*}\gamma$ are the first to be re-observed at Belle II.
The isospin asymmetry in these decays ($\Delta_{0+}$) is defined as
$$\Delta_{0+}=\frac{\Gamma(B^{0}\to K^{*0}\gamma)-\Gamma(B^{+}\to K^{*+}\gamma)}{\Gamma(B^{0}\to K^{*0}\gamma)+\Gamma(B^{+}\to K^{*+}\gamma)},$$
which constitutes a reliable observable as most of the theoretical uncertainties cancel in the ratio \cite{kstpred}.
Recent measurement~\cite{isospinV} has shown evidence for isospin violation with $3.1 \sigma$ significance, drawing lots of attention to these decays.
If this effect is real, then it can be observed with 5\,ab$^{-1}$ data at Belle II.
The current analysis~\cite{kstgammaplot} is based on following major decay channels, 
\begin{itemize}
    \item $B^{0}\to K^{*0}[\to K^{+}\pi^{-}]\gamma$,
    \item $B^{+}\to K^{*+}[\to K^{0}_{S}\pi^{+}]\gamma$, and
    \item $B^{+}\to K^{*+}[\to K^{+}\pi^{0}]\gamma$
\end{itemize}
reconstructed in a sample corresponding to 2.62 fb$^{-1}$.
The dominant background is light-quark production $e^{+}e^{-} \to q\bar{q}$, also known as continuum background.
These events have jetlike structure making them easily distinguishable from $B\bar{B}$ events, where the spatial distribution of particles is spherical.
A boosted decision tree classifier ~\cite{fastbdt}, based on several event-shape variables is trained to suppress continuum background.
The selection criterion on the classifier output is optimized by maximizing S/$\sqrt{\text{S+B}}$, where S and B are the number of signal and background events in the signal region.

Two kinematic variables called the energy difference ($\Delta E$) and beam-energy constrained mass ($M_{\textrm{bc}}$) are used for the signal $B$-meson reconstruction.
$\Delta E$ is the difference between the energy of the reconstructed $B$ meson in the center-of-mass frame and half of the collision energy. $M_{\textrm{bc}}$ is the mass of the reconstructed $B$ candidate  in the center-of-mass frame with its energy being replaced by half of the collision energy.
A tight requirement $\Delta E\in [-0.2,0.08]$ GeV is applied to suppress combinatorial background.
The signal yield is then obtained by performing an unbinned maximum-likelihood fit to the $M_{\textrm{bc}}$ distribution.
The combined significance of the above three channels exceeds $5\sigma$.
The obtained results are listed in Table~\ref{tab:K*gBR}.
  	
  	\begin{table}[!ht]
  	\centering
  	\caption{Results of the $B\to K^{*}\gamma$ analysis.}
  	\label{tab:K*gBR}
			\begin{tabular}{ |c| c |c |}
			\hline
			& Signal yield  & Significance \\ 
			& (stat. error only) &\\
			\hline 
			{ $B^{0}\to K^{*0}[K^{+}\pi^{-}]\gamma$}&$19.1 \pm 5.2$&4.4$\sigma$\\
			{ $B^{+}\to K^{*+}[K^{+}\pi^{0}]\gamma$}&$9.8\pm3.4$&3.7$\sigma$\\
			{ $B^{+}\to K^{*+}[K^{0}_{S}\pi^{+}]\gamma$}&$6.6\pm3.1$&2.1$\sigma$\\
			\hline
		\end{tabular}
	\end{table}

\subsection{Branching fraction measurement}

The theoretical predictions of the inclusive decays are more precise than the exclusive decays because they have no form factor dependence \cite{inclusive}.   
Profiting from that, the branching fraction of $\bar{B}\to X_{s}\gamma$ provides an important constraint on NP models such as extended Higgs boson sector or supersymmetry~\cite{higgs}. 
Relying on an effective theory approach, this allows to set stringent constraints on the Wilson coefficients $C_{7}$ and $C_{8}$~\cite{cpv_theory}.
In a semiinclusive analysis the hadronic system $X_{s}$ is reconstructed with several exclusive decays that contain an odd number of kaons in the final state.
We can separately measure $\bar{B}\to X_{s}\gamma$ and $\bar{B}\to X_{d}\gamma$ only in the semiinclusive method.
For the fully inclusive method, where only the hard photon is reconstructed at the signal side, the other $B$ meson is reconstructed from either hadronic or semileptonic decays. 

So far, all measurements apply a threshold on the photon energy $E_{0} = [1.7, 2.0]$\,GeV, and assumptions 
are  made to extrapolate the results down to 1.6\,GeV to be consistent with theory predictions.
This extrapolation introduces a systematic uncertainty to the result.
Another dominant source of uncertainty in the fully inclusive $\bar{B} \to X_{s} \gamma$ analysis arises from neutral hadrons faking the photon.
If the threshold value $E_{0}$ is lowered, the neutral hadron background increases causing a larger uncertainty.
So there is a trade-off between the two types of uncertainty.
Dedicated studies on the spatial distribution of the signals in the electromagnetic calorimeter at Belle II, which were not tried at Belle, can help improve these systematic uncertainties. In the hadronic tagging method S/B is very good at the cost of low signal efficiency [$\mathcal{O}(0.1\%)$].
Thanks to the large data set, the hadronic tagging analysis is possible at Belle II.
One of the dominant systematic uncertainties in the semiinclusive method is due to missing decay modes, which can be reduced at Belle II with the help of the larger data set.
The relative uncertainties in the measured branching fraction are listed in Table~\ref{tab:br}.
\begin{table}

    \caption{Expected fractional uncertainty on the Belle II measurement of BF $(\bar{B} \to X_{s} \gamma)$ 
    for each analysis technique in two scenarios of luminosity
    with $E_{0} = 1.9$\,GeV.}
    \label{tab:br}
	\begin{tabular}{ |c | c| c| c }
			\hline
			 Method  &  Belle II 5 ab$^{-1}$& Belle II 50 ab$^{-1}$\\
			\hline
		     Leptonic tag & 3.9\% & 3.2\%\\
			Hadronic tag& 7.0\%&4.2\%\\
			Semiinclusive & 7.3\% & 5.7\%\\
			\hline
	
	\end{tabular}
\end{table}

\subsection{CP violation measurement}

The time-integrated CP asymmetry for $\bar{B}\to X_{q} \gamma$ decays is defined as $$\mathcal{A}_{\text{CP}}(\bar{B} \to X_{q}\gamma) = \frac{\Gamma{(\bar{B}\to X_{q} \gamma})-({B}\to X_{\bar{q}} \gamma)}{\Gamma{(\bar{B}\to X_{q} \gamma})+({B}\to X_{\bar{q}} \gamma)}.$$
Deviation of $\mathcal{A}_{\text{CP}}(\bar{B} \to X_{s(d)} \gamma)$ from the SM prediction is a sign of NP that would modify the Wilson coefficients $C_{7}$ and $C_{8}$~\cite{cpv_theory}.
The theory uncertainties \cite{cpvxdgamma} in these observables are quite large:
\begin{eqnarray}
  \mathcal{A}^{\text{SM}}_{\text{CP}}(\bar{B}\to X_{s}\gamma) = [-0.6\%, 2.8\%] , \\
  \mathcal{A}^{\text{SM}}_{\text{CP}}(\bar{B}\to X_{d}\gamma) = [-62\%, 14\%] .
\end{eqnarray}

However, the asymmetry combined for $X_{s}$ and $X_{d}$ states is expected to be small in the SM,  $\mathcal{A}^{\text{SM}}_{\text{CP}} (\bar{B} \to X_{s+d} \gamma) = \mathcal{O}(\Lambda_{\text{QCD}/m_{b}})$  because of the CKM-matrix unitarity.
The corresponding Belle measurement $\mathcal{A}^{\text{SM}}_{\text{CP}}(\bar{B}\to X_{s+d} \gamma) = 2.2\pm3.9 \text{ (stat.)} \pm0.9{\text{ (syst.)}}$~\cite{leptonic_tag}, based on the leptonic tag, is consistent
with the SM prediction and uncertainty is dominated by the
sample size, which offers promising opportunities for the Belle II data set.
The dominant systematic uncertainty is due to the asymmetry of $B\bar{B}$ backgrounds, which are subtracted.
The estimation of this asymmetry from sidebands will be more accurate with a larger data set.
In fact, using the hadronic tag method we can precisely measure the asymmetry of both charged and neutral $\bar{B}\to X_{s}\gamma$ decays and dominant peaking backgrounds.
The Belle II data set will allow for test the assumption that
the CP violating asymmetry is independent of the $X_s$ decay mode.
The systematic uncertainty due to detector asymmetry can also be reduced using a large data set since this is also measured from sidebands or control samples.

The isospin asymmetry introduced earlier, can also be measured in the inclusive analysis of $B \to X_{s} \gamma$ decays.
Another reliable observable is the difference of direct CP asymmetries between the charged and neutral $B$ decays, $\Delta A_{\text{CP}} = A_{\text{CP}}(B^{+} \to X^{+}_{s} \gamma) - A_{\text{CP}} (B^{0} \to X^{0}_{s} \gamma)$, which can be shown to be proportional to Im($\frac{C_{8g}}{C_{7\gamma}}$)~\cite{cpvxdgamma}.
In the SM, $C_{7}$ and $C_{8}$ are both real, therefore $\Delta A_{\text{CP}}$ is zero, but in several NP models~\cite{cpvxdgamma,deltaAcp_theory2,deltaAcp_theory3} $\Delta A_{\text{CP}}$ can reach the level of $10\%$.
Since the distinction between charged and neutral $B$ decays is necessary to measure these two observables, only the semiinclusive and hadronic-tag methods can be used.
So far, measurements~\cite{deltaAcp_measure1,deltaAcp_measure2} are consistent with the SM.
In these studies statistical uncertainties dominate and can be improved at Belle II.
Another dominant uncertainty is due to the production ratio  of $B^{+} B^{-}$ and $B^{0} \bar{B}^{0}$ from the $\Upsilon(4S)$ decay ($f_{+-}/f_{00}$).
At Belle II, this factor can be measured with better precision using double semileptonic decay $\bar{B} \to D^{*} \ell^{-} \bar{\nu}$.
The expected fractional uncertainties on the Belle II measurements of the discussed asymmetries are shown in Table~\ref{tab:cpv} for each analysis technique
in two scenarios of luminosity.

\begin{table}[!ht]
    \caption{Expected uncertainties on the Belle II measurements of the CP
and isospin asymmetries for each analysis technique in two scenarios of
luminosity with $E_{0} =1.9$ GeV.}
    \label{tab:cpv}
	\begin{tabular}{|c |c | c| c| c }
			\hline
			 Observable &Method  &  Belle II & Belle II\\
			 && 5 ab$^{-1}$&50 ab$^{-1}$\\
			 \hline
			
			 $A_{\text{CP}}(B \to X_{s+d} \gamma)$ &Leptonic tag&1.5\%&0.48\%\\
			  $A_{\text{CP}}(B \to X_{s+d} \gamma)$ &Hadronic tag&2.2\%&0.70\%\\

			  $\Delta A_{\text{CP}}(B \to X_{s+d} \gamma)$ &Semiinclusive&0.98\%&0.30\%\\
			  $ \Delta A_{\text{CP}}(B \to X_{s+d} \gamma)$ &Hadronic tag&4.3\%&1.3\%\\
			 
			  $\Delta_{0+}(B \to X_{s+d} \gamma)$ &Semiinclusive&0.81\%&0.63\%\\
			  $\Delta_{0+}(B \to X_{s+d} \gamma)$ &Hadronic tag&2.6\%&0.85\%\\
			 
			\hline
	
	\end{tabular}
\end{table}

\section{Electroweak penguin $B$ decays}

\begin{figure}[h]
    \begin{tikzpicture}
	\begin{feynman}
	
	\vertex(c){ $b$};
	\vertex[right=3.5cm of c] (d){ $s$};
	\vertex[right=1cm of c] (e);
	\vertex[right=1.5cm of e] (f);
	\vertex[above=1cm of e](g);
	\vertex[above=1cm of f](h);
	\vertex at ($(g)!0.6!(h)!2cm!60:(h)$) (i){ $\ell^{+}$};
	\vertex at ($(g)!0.6!(h)!1.5cm!30:(h)$) (j){ $\ell^{-} $};
	
	\vertex [below=0.5em of e](z);
	
	\vertex [below=0.5em of f](z1);
		
	\diagram* {
		(c) --[fermion] (e),
		(e) --[fermion, edge label={ $t$}] (f),
		(f) --[fermion] (d),
		(e) -- [boson, edge label= {\small \(W^{-}\)}] (g),
		(h) -- [ boson, edge label= \(W^{-}\)] (f),
		(g) -- [ fermion, edge label=  $\nu_{\ell}$] (h),
		(i) -- [ fermion] (g),
		(h) -- [ fermion] (j),
	};
	\end{feynman}	
	\end{tikzpicture}
	\begin{tikzpicture}
	\begin{feynman}
	\vertex (b1) {$b$};
	\vertex [right=2em of b1] (b2);
	\vertex [right=4em of b2] (b3);
	\vertex [right=2em of b3] (b4) {$s$};
	\vertex at ($(b2)!0.5!(b3)!0.9cm!90:(b3)$) (g1);
	\vertex [above=6em of b3] (g2);
	\vertex [above=7em of b4] (l1) {$\ell^{+}$};
	\vertex [below=2em of l1] (l2) {$\ell^{-}$};
	
	\diagram* {
		(b1) -- [fermion] (b2) -- [fermion, edge label={{$t$}}] (b3) -- [fermion] (b4),
		(b2) -- [boson, quarter left] (g1) -- [boson, edge label = {{$W^{-}$}}, quarter left] (b3),
		(g1) -- [photon, edge label={\tiny $\gamma/Z^{0}$}] (g2),
		(l1) -- [fermion, bend right] (g2) -- [fermion, bend right] (l2),
	};
	\end{feynman}
    \end{tikzpicture}
	
   \caption{Feynman diagrams for the $b\to s\ell^{+} \ell^{-}$ process. Left diagram is known as box diagram and right diagram is known as penguin diagram.}
    \label{fig:my_label8}
	\end{figure}
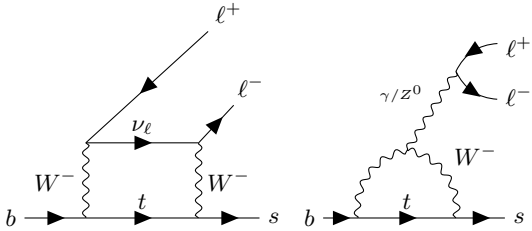

Electroweak penguin amplitudes mediate the  $b \to s \ell^{+} \ell^{-}$ process.
The dominant Feynman diagrams in the SM are shown in Fig.\ref{fig:my_label8}.

\subsection{Lepton flavor universality test }

Within the SM, gauge bosons couple equally to different flavors of lepton.
The only non-universality between leptons is their coupling with the Higgs boson as it depends on their mass, but still it has negligible effect on the BF of the decays.
Therefore, the ratios of branching fractions, referred to as $R$-ratios, $$R_{H}[q_0^{2},q_1^{2}] =  \frac{\int^{q_1^{2}}_{q_0^{2}}dq^{2}\frac{d\Gamma{(B \to H \mu^{+} \mu^{-})}}{dq^{2}}}{\int^{q_1^{2}}_{q_0^{2}}dq^{2}\frac{d\Gamma{(B \to H e^{+} e^{-})}}{dq^{2}}},$$  with $H$ being a hadron,
are expected to be unity up to corrections from the phase-space difference due to different masses, where $q^{2}$ is the dilepton invariant mass squared.
These $R$-ratios are very reliable observables, as the theoretical uncertainties from CKM factors, form factors and other hadronic effects cancel since they are common in the numerator and denominator.
The dilepton mass ranges corresponding to charmonium resonances are vetoed.
This leads to two dilepton square-mass regions, namely low-$q^{2}$ ($q^{2}\in[1,6]$ GeV$^{2}/c^{2}$) and high-$q^{2}$ ($q^{2} > 14.4$ GeV$^{2}/c^{2}$) regions. Within these two regions the SM predictions for $R$ ratios are 1 with high precision.
For example, $R^{\text{SM}}_{K}[1,6] = 1.000\pm 0.001$~\cite{smrk}.

The main experimental challenge is understanding the difference in reconstructed efficiency between electron and muons. The most important difference is introduced by the bremsstrahlung process, which causes electrons to radiate a significant amount of energy.
So far, LHCb provided the most precise measurement of both $R_{K^{(*)}}$ in the low-$q^{2}$ region ~\cite{rklhcb,rkstlhcb}.
The $R_{K^{(*)}}$ measurement result is compatible with the SM at the level of 2.6 (2.5) standard deviations. 
    
    \begin{figure}[h]
		\includegraphics[width=7cm]{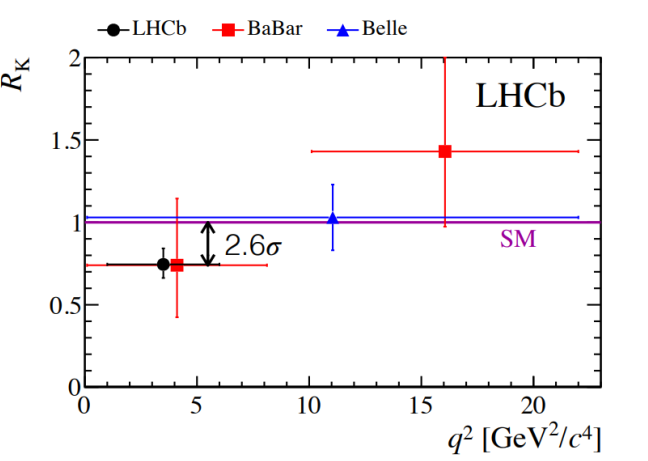}
		\includegraphics[width=7cm]{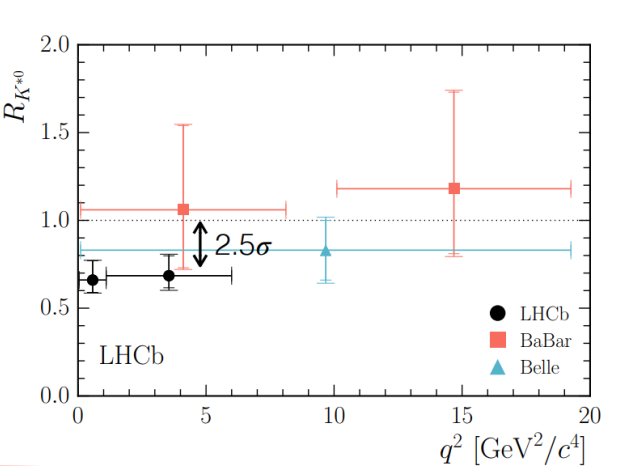}

    \caption{Measurement of $R_{K}$ (top) and $R_{K^{*}}$ (bottom) in different experiments.}
    \label{fig:my_labelrk}
    \end{figure}

A previous measurement by Belle~\cite{rkbelle,rkbelle1} has higher uncertainty, and is consistent with both SM and LHCb measurement.
Belle already measured $R$-ratios in the high $q^{2}$ bins, while no measurement from LHCb has been reported ~\cite{rkbelle}.  At Belle II electron and muon modes has almost similar reconstruction efficiency.
Using a larger data set, the future Belle II measurement can shed light on these  $R_{K^{(*)}}$ anomalies.
If the $R_{K}$ anomaly is real due to NP, we should be able to establish it with $5\sigma$ significance using around 20\,ab$^{-1}$ of Belle II data.
Thanks to the clean environment, Belle II can also study inclusive $B \to X_{s} \ell^{+} \ell^{-}$ decay and measure $R_{X_{s}}$.
Furthermore, Belle II can measure individually charged and neutral channels in $B \to K^{*0/+} \ell \ell$.
In Table~\ref{tab:lfv} expected resolutions of  $R$-ratio observables are listed.

\begin{table}[!ht]

    \caption{Expected resolutions on the observables that test lepton flavor universality at Belle II.}
    \label{tab:lfv}
	\begin{tabular}{|c |c | c| c }
			\hline
			 Observable &  Belle II 5 ab$^{-1}$& Belle II 50 ab$^{-1}$\\
			 \hline
			 $R_{K}$[1,6] GeV$^{2}/c^{2}$ &11\% & 3.6\%  \\
			 $R_{K}$[$>$14.4] GeV$^{2}/c^{2}$ &12\% & 3.6\%  \\
			 $R_{K^{*}}$[1,6] GeV$^{2}/c^{2}$ &10\% & 3.2\%  \\
			 $R_{K^{*}}$[$>$14.4] GeV$^{2}/c^{2}$ &9.2\% & 2.8\%  \\
			 $R_{X_{s}}$[1,6] GeV$^{2}/c^{2}$ &12\% & 4.0\%  \\
			 $R_{X_{s}}$[$>$14.4] GeV$^{2}/c^{2}$ &11\% & 3.4\%  \\

			\hline
	
	\end{tabular}
\end{table}

\subsection{Angular analysis of $B \to K^{*} \ell^{+} \ell^{-}$}

\begin{figure}[h]
		\includegraphics[width=5.5cm]{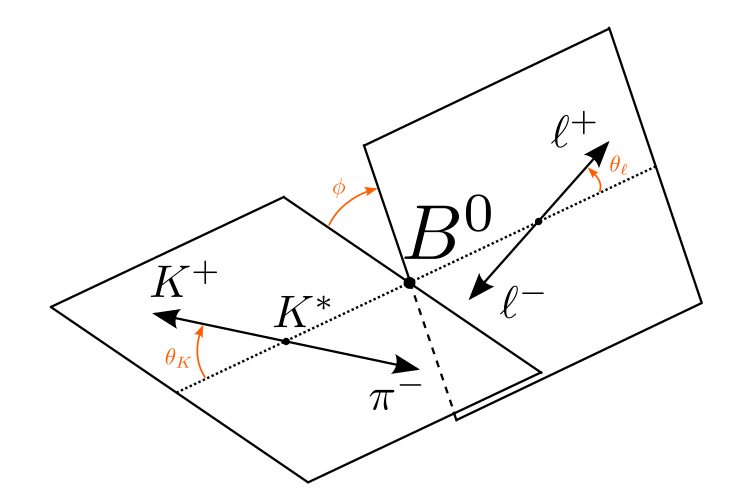}
    \caption{ Definitions of angles in the $B^{0} \to K^{*0} \ell \ell$ decay}
    \label{fig:my_label4}
\end{figure}

An angular analysis of $B \to K^{*}[K\pi] \ell^{+} \ell^{-}$ decays provides several observables that are sensitive to NP.
The angular distributions are completely described by four independent kinematic variables, chosen as $q^{2} = M^{2}_{\ell^{+} \ell^{-}}$ and three angles $\cos \theta_{\ell}$, $\cos \theta_{K}$, and $\phi$.
The angle $\theta_{\ell}$ is the angle between the $\ell^{+} (\ell^{-})$ momentum and the momentum of dilepton system in the $B (\bar{B})$ rest frame.
The angle $\theta_{K}$ is the angle between the direction of kaon  and the $K^{*}$ momentum in the $B (\bar{B})$ rest frame.
The angle $\phi$  is the angle between the decay plane of $\ell^{+} \ell^{-}$ and  $K^{*}$.
These angles are described in Fig.~\ref{fig:my_label4}.
 
The differential decay rate in terms of angular variables is given by,
$$\frac{d^{4}\Gamma(\bar{B} \to \bar{K^{*}} \ell^{+} \ell^{-})}{d \cos{\theta_{\ell} d \cos{\theta_{K}d\phi dq^{2}}}} = \frac{9}{32\pi} \sum_{j} I_{j} f_{j}(\cos{\theta_{\ell}},\cos{\theta_{K}},\phi),$$
$$\frac{d^{4}\Gamma({B} \to {K^{*}} \ell^{+} \ell^{-})}{d \cos{\theta_{\ell} d \cos{\theta_{K}d\phi dq^{2}}}} = \frac{9}{32\pi} \sum_{j} \bar{I}_{j} f_{j}(\cos{\theta_{\ell}},\cos{\theta_{K}},\phi),$$
where $I_{j}$ and $\bar{I}_{j}$ are functions of $q^{2}$ and depend on the $K^{*}$ transversity amplitude ~\cite{angular}.
The angular dependence of each term comes from $ f_{j}(\cos{\theta_{\ell}},\cos{\theta_{K}},\phi)$, parametrized with spherical harmonics associated with different polarisation states of the $K^{*}$ and dilepton system.
The self-tagging nature of the $B \to K^{*} \ell^{+} \ell^{-}$ decay allows for determining both CP-averaged and CP-asymmetric quantities that depends on the coefficients, 
$$S_{i} = (I_{i}+\bar{I}_{i})/\frac{d\Gamma}{dq^{2}},$$
$$A_{i} = (I_{i}-\bar{I}_{i})/\frac{d\Gamma}{dq^{2}}.$$

It is possible to exploit symmetry relations to construct observables that are free from form-factor uncertainties at leading order in the 1/$m_{b}$ expansion~\cite{angular1}.
It is also possible to build reliable observables at low-$q^{2}$ exploiting the form-factor cancellation.
This includes the so-called $P^{\prime}$ series of observables \cite{angular2} defined as,
$P^{\prime}_{4} =  \frac{S_{4}}{2\sqrt{-S_{2c}S_{2s}}}$, $P^{\prime}_{5} = \frac{S_{5}}{2\sqrt{-S_{2c}S_{2s}}} $, $P^{\prime}_{6} =  \frac{S_{7}}{2\sqrt{-S_{2c}S_{2s}}}$, $P^{\prime}_{8} = \frac{S_{8}}{2\sqrt{-S_{2c}S_{2s}}} $. 
The LHCb  reported a tension in the $P^{\prime}_{5}$ observable from the $B^{0} \to K^{*0} \mu^{+} \mu^{-}$ decay  \cite{angularlhcb}.
Belle also performed the angular analysis ~\cite{angularbelle}, using its full data set with both charged and neutral $B$ mesons.
A $2.6\sigma$ tension was observed in $P^{\prime}_{5}$ of the muon modes in the region 4 GeV$^{2}/c^{2} < q^{2}<$ 8 GeV$^{2}/c^{2}$, which is the same region LHCb  observed the $P^{\prime}_{5}$ anomaly.
A lepton-flavor-dependent measurement of $P^{\prime}_{5}$ can lead to another observable $Q^{\prime}_{5} = P^{\prime \mu}_{5} - P^{\prime e}_{5}$.
There are no significant deviation from SM observed in the Belle measurement of $Q^{\prime}_{5}$.

\begin{figure}[h]
		\includegraphics[width=9.0cm]{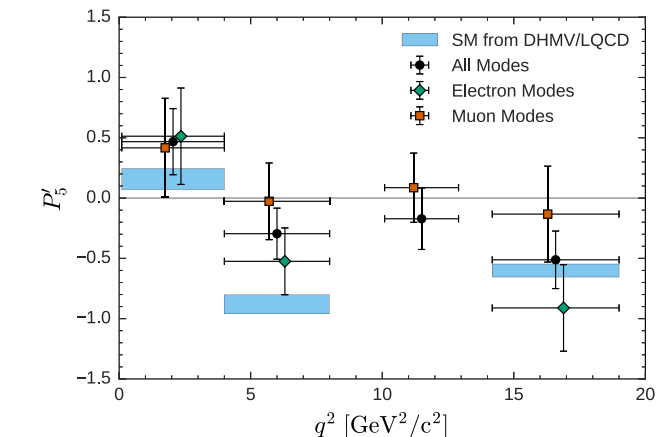}
		\includegraphics[width=9.0cm]{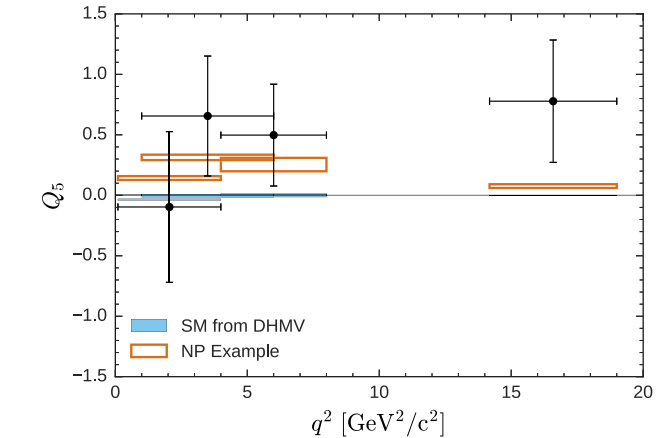}
    \caption{Measurement of $P_{5}^{\prime}$ (top) and $Q_{5}^{\prime}$ (bottom) at Belle.}
    \label{fig:my_label5}
\end{figure}

At Belle II,  the uncertainty due to peaking background can be reduced by including individual components in the fitting model as these components can be more reliably modeled with a larger data set.
The uncertainty in $P^{\prime}_{5}$, for $q^{2} \in [4,6]$ GeV$^{2}/c^{2}$ with 2.8~ab$^{-1}$ of Belle II data based on both electron and muon modes will be comparable to the 3\,fb$^{-1}$ data result of LHCb that uses the muon modes only.
A naive extrapolation  leads to the conclusion that the accuracy that can be achieved on the optimised observables at Belle II with 50\,ab$^{-1}$ is just 20\% lower than the precision that LHCb is expected to reach with 50\,$\text{fb}^{-1}$ of data\,\cite{belle2physics}.

\subsection{Missing energy channel: $B \to K^{(*)} \nu \bar{\nu}$}
\begin{figure}[h]
		\includegraphics[width=5.5cm]{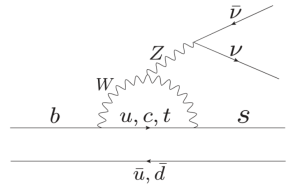}
    \caption{Electroweak penguin diagram for $b\to s\nu\bar{\nu}$.}
    \label{fig:my_label1}
\end{figure}

\begin{figure}[h]
		\includegraphics[width=5.5cm]{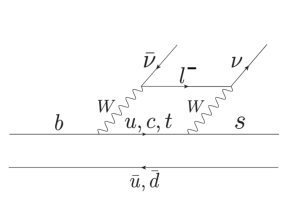}
    \caption{Box diagram for $b\to s\nu\bar{\nu}$.}
    \label{fig:my_label3}
\end{figure}

\begin{figure}[h]
		\includegraphics[width=5.5cm]{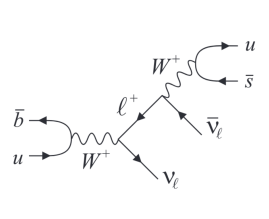}
    \caption{Tree-level diagram involving two bosons.}
    \label{fig:my_label2}
\end{figure}

The semileptonic decays mediated by $b\to s\nu\bar{\nu}$ are forbidden at tree level involving a single boson exchange.
They occur via higher-order electroweak penguin (Fig.~\ref{fig:my_label1}), box diagram (Fig.~\ref{fig:my_label3}), or tree-level transition involving at least two $W$/$Z$ bosons (Fig.~\ref{fig:my_label2}).

An advantage of the $b\to s\nu\bar{\nu}$ transition compared to $b\to s\ell^{+}\ell^{-}$ is the absence of photon mediated diagrams that lead to a pair of charged leptons.
As a consequence, the factorisation of the hadronic and leptonic current is exact for $B\to K^{(*)}\nu\bar{\nu}$ decays, which makes theoretical predictions more accurate.
Measurements of the $B\to K^{(*)}\nu\bar{\nu}$ decay rates would in principle allow the extraction of the $B\to K^{(*)}$ form factors to high accuracy.
$B$ decays involving exotic final states, e.g. dark matter candidates, are closely related to this kind of signals since the missing energy signatures within the detector is the same.
One more observable which is sensitive to NP is the $K^{*}$ longitudinal polarisation fraction ($F_{L}$) in $B \to K^{*} \nu \bar{\nu}$. An angular analysis of the 
$B \to K^{*} (\to K \pi ) \nu \bar{\nu}$
decay would allow to access the $K^{*}$ longitudinal polarisation fraction $F_{L}$.
Ref.~\cite{newst1} predicts $F^{\text{SM}}_{L} =0.47\pm 0.03$.
Another study~\cite{newst} shows that the NP-sensitive operator $\mathcal{O}_{R} =  \frac{e^{2}}{16 \pi^{2}}(\bar{s}\gamma_{\mu}P_{R}b)(\bar{\nu}\gamma_{\mu}(1-\gamma_{5})\nu)$, where $P_{R}$ is right-handed chiral projection operator, in the product expansion impacts $F_{L}$.
In other words, this observable is sensitive to right-handed quark current.

In the SM, the branching fractions of $B\to K^{+}\nu\bar{\nu}$ and $K^{*}\nu\bar{\nu}$ are $(4.0\pm 0.5)\times 10^{-6}$ and $(9.2\pm 1.0)\times 10^{-6}$, respectively \cite{knu}.
None of the decays have been observed.
They are expected to be observed with first 10\,ab$^{-1}$ of Belle II data.
A  larger data sample is required to measure $F_{L}$. 
Studies based on simplified simulated experiments show that the uncertainty on $F_{L}$ will be 0.11 with 50\,ab$^{-1}$.
The expected resolutions on the observables are listed in Table \ref{tab:nunu}.

\begin{table}

    \caption{Expected resolutions on the observables for decays mediated by $b \to s \nu \bar{\nu}$.}
    \label{tab:nunu}
	\begin{tabular}{|c |c | c| c }
			\hline
			 Observable &  Belle II 5 ab$^{-1}$& Belle II 50 ab$^{-1}$\\
			 \hline
			 BF($B^{+} \to K^{+} \nu \bar{\nu}$) &30\%&11\%\\
			 BF($B^{0} \to K^{*0} \nu \bar{\nu}$) &26\%&9.6\%\\
			 BF($B^{+} \to K^{*+} \nu \bar{\nu}$) &25\%&9.3\%\\
			 $F_{L}$($B^{0} \to K^{*0} \nu \bar{\nu}$) &--&0.079\\
			 $F_{L}$($B^{+} \to K^{*+} \nu \bar{\nu}$) &--&0.077\\
			\hline
	\end{tabular}
\end{table}

\section{Summary}

The low-background environment with constrained collision kinematics at Belle II grants access to several unique observables in rare $B$ decays.
Starting with the $B\to K^{*}\gamma$ decay, Belle II is on its way to  re-observe signal of other suppressed penguin decays.
We expect to provide strong model-independent constraints on new physics, thanks to the large data sample that will be collected by Belle II. 

\end{document}